\documentclass[preorubt,3p,times]{elsarticle}
\usepackage{amssymb}
\usepackage{hyperref}
\usepackage{float}

\journal{Nuclear Instrumentations and Methods in Physics Research, Section A}

\begin{document}

\begin{frontmatter}

\title{Multi-column Compton Camera of stacked Si pixel sensors for sub-degree angular resolution}

\author{Yasushi Fukazawa}

\affiliation{organization={Department of physics, Hiroshima University},
            addressline={1-3-1 Kagamiyama}, 
            city={Higashi-Hiroshima},
            postcode={739-8526}, 
            state={Hiroshima},
            country={Japan}}

\begin{abstract}

The Compton camera is a sensitive imaging detector for soft gamma-rays. 
Compton Reconstruction can not only give imaging capability but also remove background
events to achieve good sensitivity. However, the angular resolution is
in principle limited to several degrees.
In this paper, we propose a novel concept of Compton camera incorporating shadow effects.
We consider multi-column Compton camera (MCCC), consisting of stacked Si pixel sensors. Each of columns is separated each other to create shadow effects.
This design achieves an angular resolution of less than 1 degree within around 1 degree from the center of the field-of-view by just modifying a conventional Si-stacked Compton camera and keeping advantages (wide field-of-view and good sensitivity) of conventional Compton camera. Here we validated the 
concept of proposed Compton camera through Monte-Carlo simulation.
MCCC with 1 m column height, 0.5 mm pixel size, 100 layers, and 10 columns for the 1-D direction can distinguish two sources separated by  0.1 degree with 0.6M Compton-reconstructed events.
\end{abstract}



\begin{keyword}
Compton camera, silicon pixel sensor
\end{keyword}

\end{frontmatter}




\section{Introduction}

Gamma-ray imaging in the 100 keV to several MeV is important for astrophysical observations, medical diagnostics, and environmental radiation measurements. So far, mainly two imaging methods are conducted. One Involves a mask at the front of gamma-ray detectors. A coded-mask is often used for strophysical observations. It provides angular resolution of several arcminute (INTEGRAL \citep{INTEGRAL}, Swift/BAT \citep{BAT}). However, these detectors are simple and thus bave insufficient background rejection (rejection of background events caused by non-gamma-ray signals), detecting only bright sources. The situation is similar for medical and environmental radiation measurements; a weak gamma-ray sources cannot be detected. The other method uses Compton reconstruction. In this method, gamma-rays are scatter several times in the detector. Using deposit energy and position of each scattering or absorption points, the incident direction and energy of gamma-rays can be reconstructed. Compton reconstruction also rejects background events, so a weak source can be detected. In astrophysical observations, CGRO/OSSE \citep{COMPTEL}, Hitomi/SGD \citep{SGD}, and COSI \citep{COSI} are an example of this detector. However, angular resolution is limited to several degrees due to Doppler broadening; uncertainty of internal scattered electrons. A combination of the coded-mask and Compton reconstruction has been proposed \citep{GECCO}. Possible disadvantage of this method is that the coded-mask limits a wide field-of-view which is important for MeV gamma-ray sky survey and the coded-mask will be a background sources and thus reduce sensitivity in the two reasons. One is the case that some incident gamma-rays first  scatter in the coded-mask and then are detected in the Compton camera.  In this case, the coded-mask is passive and thus Compton reconstruction does not work and thus such events become background events. The coded-mask is usually made of lead or tungsten, and thus the scattering probability is large. The other case is that the coded-mask becomes radio-activated by protons in the satellite orbit and thus it emits gamma-ray background events. A large distance between the coded-mask and detector could also imposes  design constraints. In this paper, we propose a new concept for gamma-ray imaging with good sensitivity.

\section{Concept}

\begin{figure}[H]
 \includegraphics[scale=0.6]{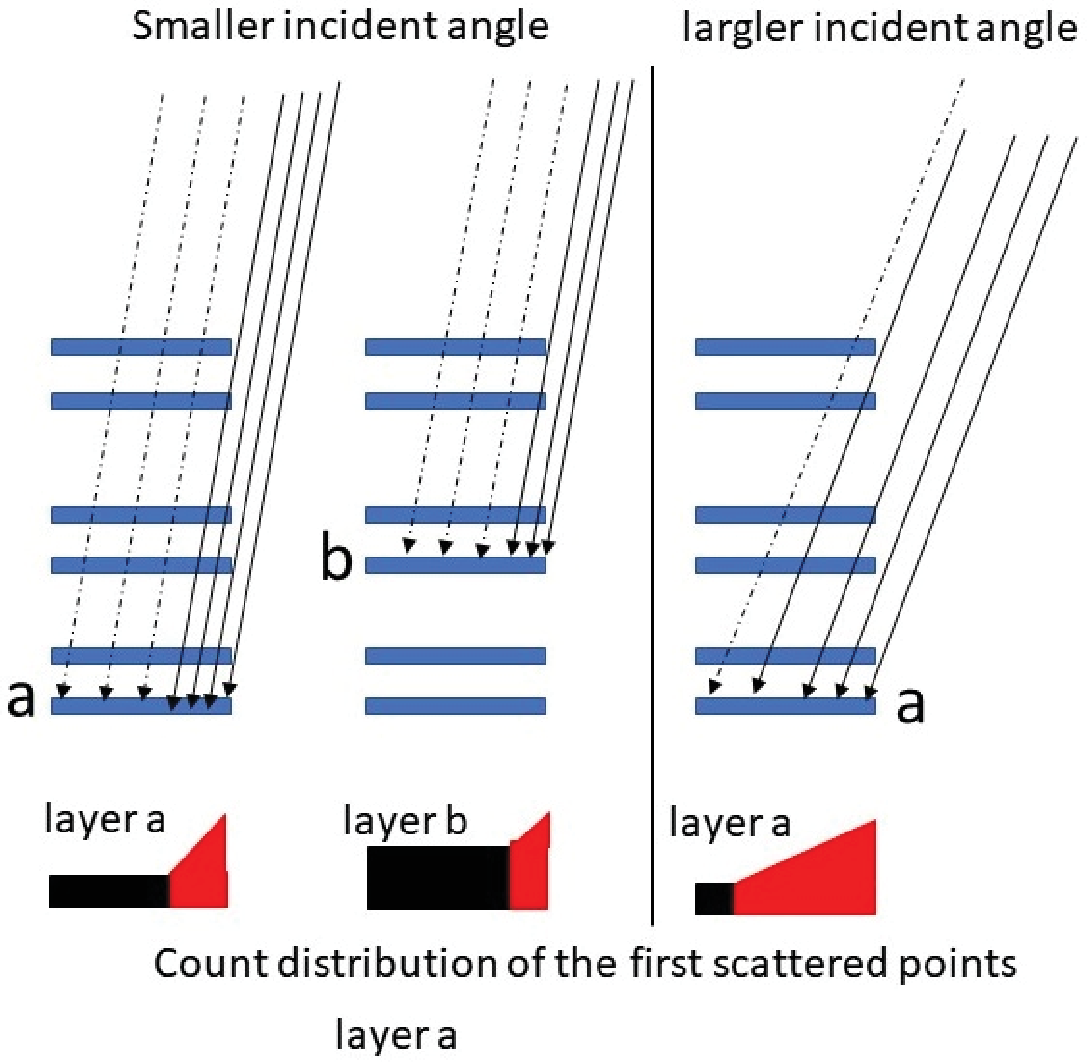}
\vspace*{3cm}
 \caption{Concept of Multi-Column Compton Camera. Blue layers represent stacked Si sensors. Arrows represent incident gamma-rays; solid arrows are gamma-rays passing through a part of layers or none above a scattering layer, and dashed arrows are gamma-rays passing through all layers above a scattering layer. Red and black at the bottom of layers represent a count distribution of 1st scattering position of a specific layer; red is a count distribution for gamma-rays passing through a part of layers or none above a scattering layer, and black is that for gamma-rays passing through all layers above a scattering layer.}
 \label{concept}
\end{figure}

The new concept is based on the shadow effect as a coded-mask, but the shadowing material is not a mask but detector parts of Compton camera with stacked semiconductor sensors. Figure \ref{concept} shows a schematic of the proposed Compton camera. The camera consists of column arrays of stacked semiconductor sensors. In the case of conventional Compton camera, each column is located with as a small separation as possible. In our design, we propose significant separation allowing incident gamma-rays to enter the region between the columns. We assume the gamma-ray source at the infinite distance, and thus the incident gamma-rays arrive nearly uniformly and in parallel.

When gamma-rays come in parallel to the normal direction of the layers, gamma-rays which enter the top of the column pass through some layers and scatter. The scattering points are uniformly distributed in each layer with more numerous scattering points in upper layers. On the other hand, when gamma-rays enter with an inclination angle against the normal direction, some gamma-rays directly enter a lower layer and scatter at the edge of the sensor.
Some incident gamma-rays pass through only a part of layers without passing through upper layers and they also scatter at the edge of the layer. These gamma-rays generate an excess count at the edge of the sensor.
The count excess closer to the
sensor edge is due to gamma-rays penetrating a smaller number of upper
layers while that closer to the excess terminal point is due to
gamma-rays penetrating a larger number of layers. The number of penetrated layers
increases for a larger distance from the sensor edge,
leading to an exponential (or almost linear) decrease in excess count from the edge.
The excess terminal point depends on the layer position; more distant from the edge for a lower layer.
The slope of the excess count depends on the incident angle; a flatter slope
for a large incident angle. In other words, we can estimate an
incident angle from the slope pattern of excess count. 
The resolution can be smaller than the angle determined by the layer
separation and pixel size due to gamma-rays passing through a part of layers. 
In contrast, for optical photons, the resolution is limited to 
the angle determined by the layer separation and pixel size because
they cannot penetrate the Si layers and thus the excess edge
count at the edge is due to photons without passing through any upper layers.

For example, in the case that the position resolution of each layer is 0.5 mm and the height of column is 1 m, we could distinguish a scale around 5/10000 rad = 0.03 degree = 1.8 arcmin by measuring the count distribution of each layer if photon statistics is high. Gamma-rays with an energy of several 0.1 to several MeV will scatter in multiple positions in the column array, but the first scattering point can be identified by Compton reconstruction. Due to poor photon statistics of counts in each layer, as many layers and columns as possible are needed. Hereafter, we call such system as multi-column Compton camera (MCCC).

Pixel sensors generally have an insensitive edge region of 0.5--1 mm, and thus gamma-rays with a small incident angle of $<$0.05 degree cannot produde excess counts at the edge region.  Also, the measurement of the incident angle with the excess edge count is not available for the incident angle larger than a few degree because the neighboring column blocks the gamma-rays and thus edge excess cannot appear.

  Note that MCCC can keep the performance of conventonal Compton cameras where columns are not separated. In the case of MCCC, all events can be analyzed by Compton reconstruction and the incident gamma-ray direction can be determined with several degrees of accuracy with rejecting background events. Accordingly, the MCCC works as a conventional Compton camera. In addition, for reconstructed events which are localized within $\sim$1 degree from the center of the field-of-view, we can apply the MCCC concept. In other words, compared to the coded-mask Compton camera, a good angular resolution is only $\sim$1 degree from the center of the field-of-view, but a wide field-of-view with good sensitivity can be kept. Strictly speaking, the separation of columns will somewhat increase gamma-ray events which do not give enough hits in the camera, reducing the reconstruction efficiency a little. However, this is a trade-off with MCCC concept. MCCC concept needs to accumulate enough number of events and thus only a bright gamma-ray sources can be used for this concept. This will be discussed in section 4. The MCCC offers a choice of MeV observation detector design and the coded-mask or MCCC can be chosen according to the mission of the MeV gamma-ray observatory.

The MCCC requires many sensors with moderate position resolution. Si double-sided strip sensors(DSSDs) have been often used for scatterer. DSSDs require external signal processing circuits that consume significant power and the external circuit is not preferred since it becomes a passive area for Compton camera. Conventional CMOS sensors have good position resolution but the power consumption is too high to be arrayed. On the other hand, one of the HV-CMOS sensors, such as AstroPix \citep{AstroPix,AstroPix2}, are being developed. This has a moderate position resolution of 0.25--1.0 mm. Supplying 400--500 V bias, full depletion of 0.5 mm thickness can be available, leading to being sensitive to gamma-ray. The target power consumption is $<1$ mW/cm$^2$. Multiple sensors can be read out via daisy-chain. AstroPix will be used in the MeV gamma-ray satellite project AMEGO-X \citep{AMEGO-X} and the demonstration to integrate sensors is planned as a Compair-2 project around 2026--2027. AstroPix will also be used for the electron-ion collider (eIC) experiment at Brookhaven National Laboratory \citep{eIC}. These projects will use tens of thousands of AstroPix sensors. Therefore, AstroPix is a good candidate for the MCCC. Especially, the AMEGO-X detector consists of 2x2 towers, each of which contains 40 layers of AstroPix sensors, and thus a modification of such tower design could realize the MCCC concept.

\section{Monte-Carlo Simulation}

Monte-Carlo simulation was performed to verify the concept of MCCC with MEGAlib\citep{MEGAlib,MEGAlib2}. MEGAlib is based on the Geant 4 simulator library and has been used for detector simulation in gamma-ray astronomy, such as COSI and AMEGO-X \citep{AMEGO-X}. MEGAlib provides a user-friendly interface for the detector geometry setup and radiation input. In addition, it offers Compton reconstruction \citet{Recon}. It is based on the examination of the structure of the event. This includes the relative locations of the hits, sequencing of the hits within the event, and detector properties between the hit locations. Several algorithms (correlation-based algorithms, classic Compton sequence reconstruction methods, and Bayesian reconstruction approaches) are employed.

\begin{figure}[H]
 \includegraphics[scale=0.4]{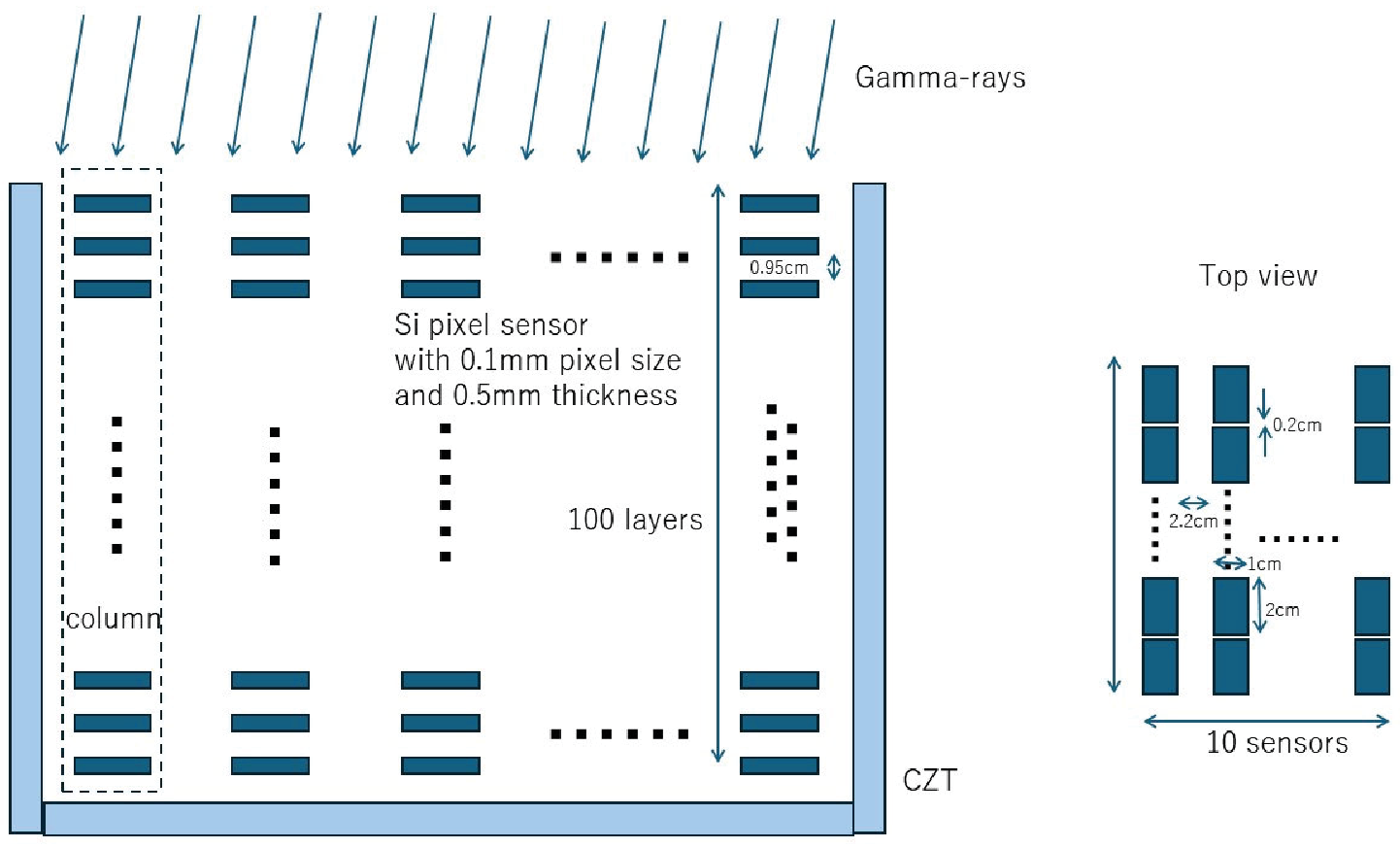}
\vspace*{1.5cm}
 \caption{Detector geometry for simulation.}
 \label{simsetup}
\end{figure}

Here we studied the one-dimensional angular resolution. The MCCC configuration was modified from that of AMEGO-X which has four towers, each of which consists of 40 layers of 40$\times$40 array of 2$\times$2 cm$^2$ pixel sensors\citep{AMEGO-X}. In the simulation, we constructed MCCC as shown in figure \ref{simsetup}, by stacking pixel sensors with a pixel size of 0.5 mm, a sensitive area of 2$\times$4 cm$^2$, a thickness of 0.5 mm, an energy resolution of 5 keV, a threshold of 25 keV, and an insensitive edge region of 0.5 mm. MCCC consists of 10$\times$10 columns, and one column is a stack of 100 layers of pixel sensors with 2$\times$4 cm$^2$ area. Each layer is separated by 2.2, 0.2, and 0.95 cm for the X, Y, and Z directions, respectively. This MCCC is sensitive only to the X direction.  CZT sensors with 2-cm thickness surround the 10$\times$10 column array on four sides and the bottom to enhance the Compton reconstruction efficiency. The energy and position resolution of the CZT at 600 keV are 3 keV and 0.2 mm, respectively. A parallel beam of 600 keV gamma-ray photons is injected from the top of the MCCC with a certain inclination angle with respect to the normal direction of the MCCC layers.

\begin{figure}[H]
  \begin{minipage}{0.5\hsize}
    (a)\\  \vspace*{0.6cm}
   \centering
   \includegraphics[scale=0.4]{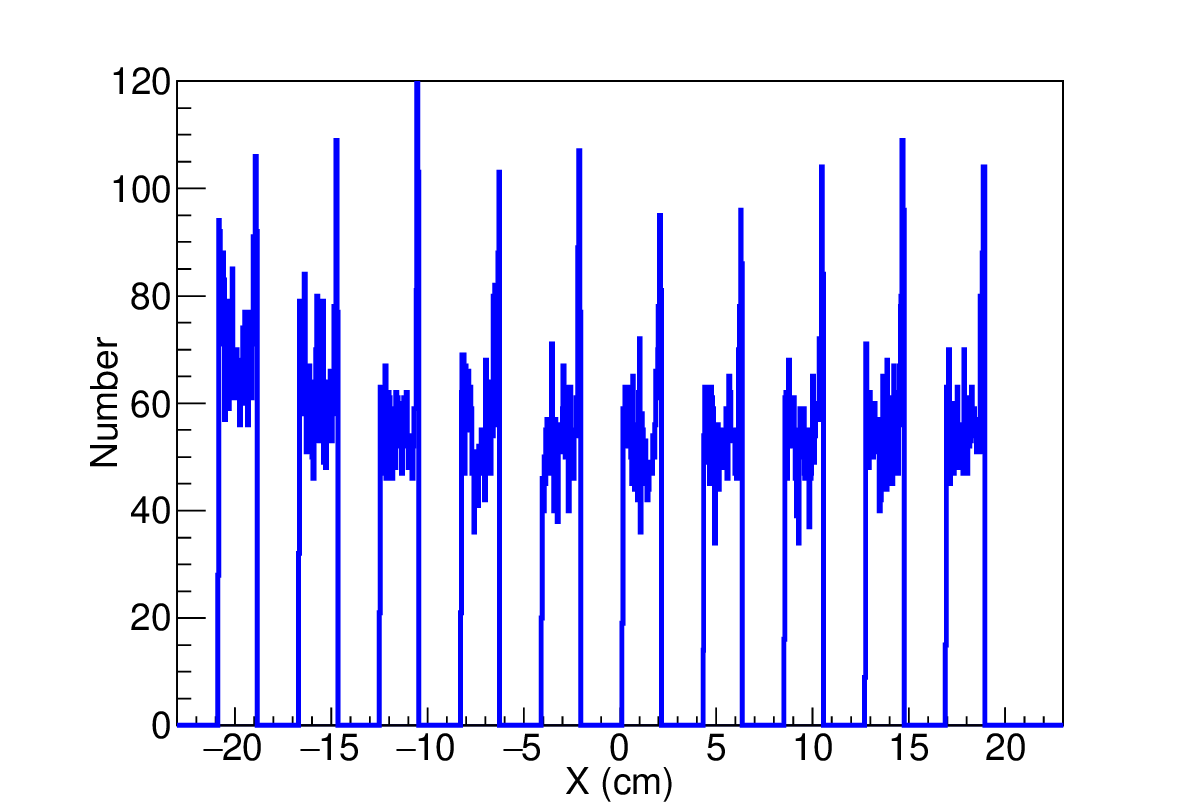}
   \end{minipage}\quad
   \begin{minipage}{0.6\hsize}
    (b)\\  \vspace*{0.6cm}
   \centering
   \includegraphics[scale=0.4]{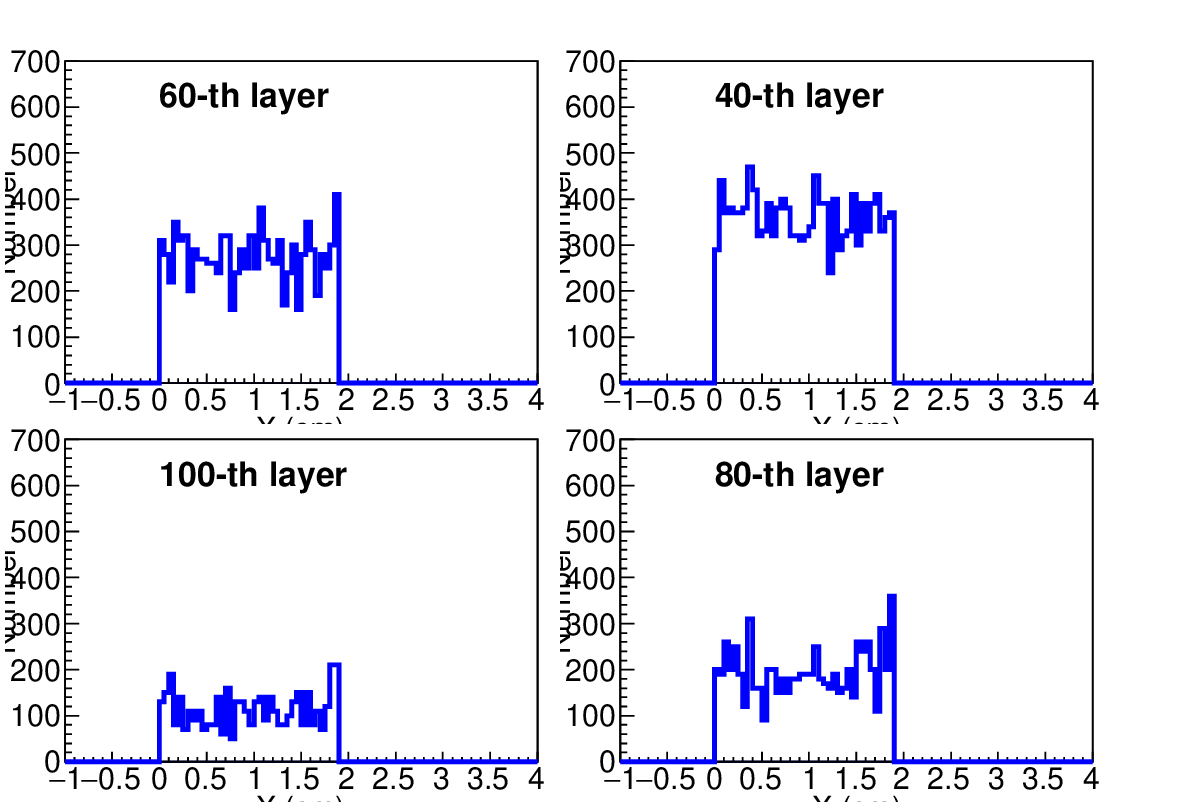}
   \end{minipage}
   \caption{An example of count distribution of the first Compton scattering position in MCCC in the case of incident angle of 0.14 degree. (a) One for the 100th layer.  (b) Count distribution summed over 10 columns in each layer. From top-right to bottom-left is one for 40, 60, 80, 100th layer.}
   \label{countdist}
\end{figure}

Figure \ref{countdist}(a) presents an example of the number distribution of the first scattering points at the bottommost (100th) Si layers for the 600 keV gamma-ray injection with an inclination of 0.14 degree. The total number of Compton-reconstructed photons is 6.3M. The horizontal axis is the coordinate in units of cm for the X direction. There are 10 sensitive regions in each layer, corresponding to the sensors in the 10 columns. The right sharp peak in each sensor represents the events that passed through a part of the upper layers. This peak can be used for estimating incoming direction. Figure \ref{countdist}(b) shows the count distribution of the selected layers. In this plot, the count histograms of each column on the same layer are summed over 10 columns to increase the statistics. The flat component over one layer decreases toward the lower layer. This flat component is caused by gamma-rays penetrating through the upper layers as described in section 2. The right-edge component appears clearly in the lower layer. Histograms of all 100 layers  can be used to determine the direction of the gamma-ray sources. Various algorithms can be considered to determine the gamma-ray incident direction. Hoeever, in this study, we determined the direction by simply fitting the count distribution.

\begin{figure}[H]
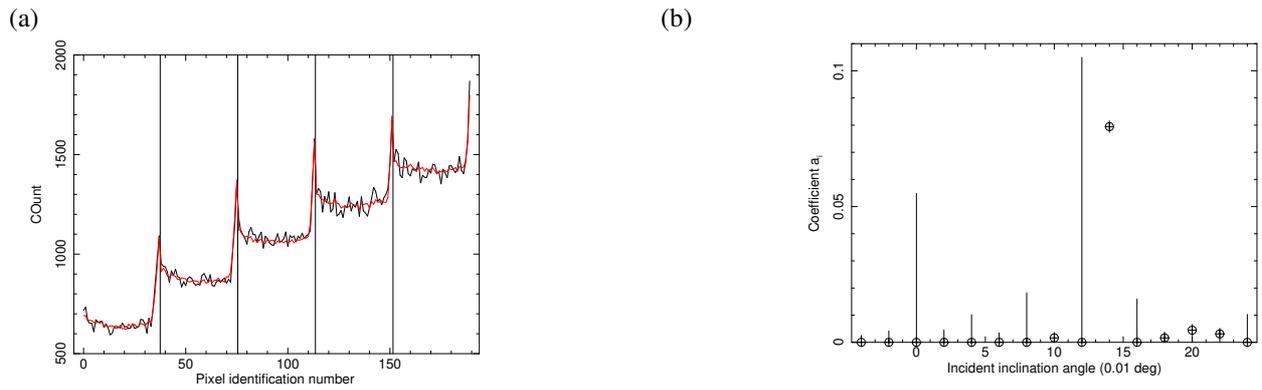

  \begin{minipage}{0.5\hsize}
    (a)
   \end{minipage}\quad
   \begin{minipage}{0.6\hsize}
    (b)
   \end{minipage}
   \begin{minipage}{0.6\hsize}
   \includegraphics[scale=0.25,angle=-90]{a120p14.eps}
   \end{minipage}\quad
   \begin{minipage}{0.6\hsize}
   \includegraphics[scale=0.25,angle=-90]{b120p14.eps}
   \end{minipage}
 \caption{An example of fitting of the count distribution for the first Compton scattering position to obtain the incident angle. Here the incident gamma-rays are injected with an inclination angle of 0.14 degree.  (a) Count distribution of data (black) and a model function (red) against the pixel identification number $k$. From left to right, the count distribution for the sum of 91-100th, 81-90th, 71-80th, 61-70th, and 51-60th layers, whose borders are indicated by the vertical lines.  (b) The best-fit coefficients $a_i$ for each incident angle model (see text in detail), obtained by fitting. The horizontal axis is an incident inclination angle in unit of 0.01 degree.}
   \label{fitting1}
\end{figure}

  To determine the incident angle, we analyzed the excess count profile at the sensor edge. We modeled the excess count profile with the simulated profile of various incident angles and determined which incident angle best fit the data. In order to generate the model for fitting, we prepared a set of histograms as shown in figure \ref{fitting1} for injection angles of -0.5 to 0.5 degree with a step of 0.02 degree. Approximately 6.3M reconstructed events were used to generate a count distribution for each angle. The histograms of each layer were summed over 10 continuous layers and then 10 histograms were generated. Since the edge peak was clear for the lower layer, five histograms generated from lower 51-100 layers were used for modeling. Each of histogram has 38 bins for the sensitive region. These five histograms were combined into one histogram. The histogram $M_i(x_k)$, where $i$ is the injection angle identification number and $k$ is the pixel identification number, can be an angular response of MCCC. Then, we generated the same histogram $C(x_k)$ for a particular incident angle, and subsequently it was fitted with a linear combination of $M_i(x_k)$ as $ C(x_k)=\sum_i a_i M_i(x_k)$.

  Figure \ref{fitting1}(a) shows an example of fitting, where the fitted profile is generated from 0.63M reconstruction events for the incident angle of 0.14 degree. The combined 5 histograms are shown, together with the model. In addition to the flat component which is caused by gamma-rays penetrating the upper layers, excess count profiles are seen at the right edge of each histogram.
Figure \ref{fitting1}(b) plots the coefficient $a_i$ as a function of $i$, corresponding to a unit of 0.02 degree. This $a_i$ value becomes 1 when the number of incident gamma-rays is the same as that in the model (6.3M). 
A large value appears at 14, indicating that the incident angle is likely 0.14 degree. The incident angle cannot be determined for the incident angle of $<$0.06 degree, due to the insensitive edge region. We performed this fitting for 20 sets of 0.315M events for 0.1 degree incident angle, took a coefficient-weighted mean position where we ignored values of the positions at three bins away from the mean position. Then, we determined the mean and variance of 20 obtained mean position, and they become 0.1010 degree and 0.0028 degree, respectively.

\begin{figure}[H]
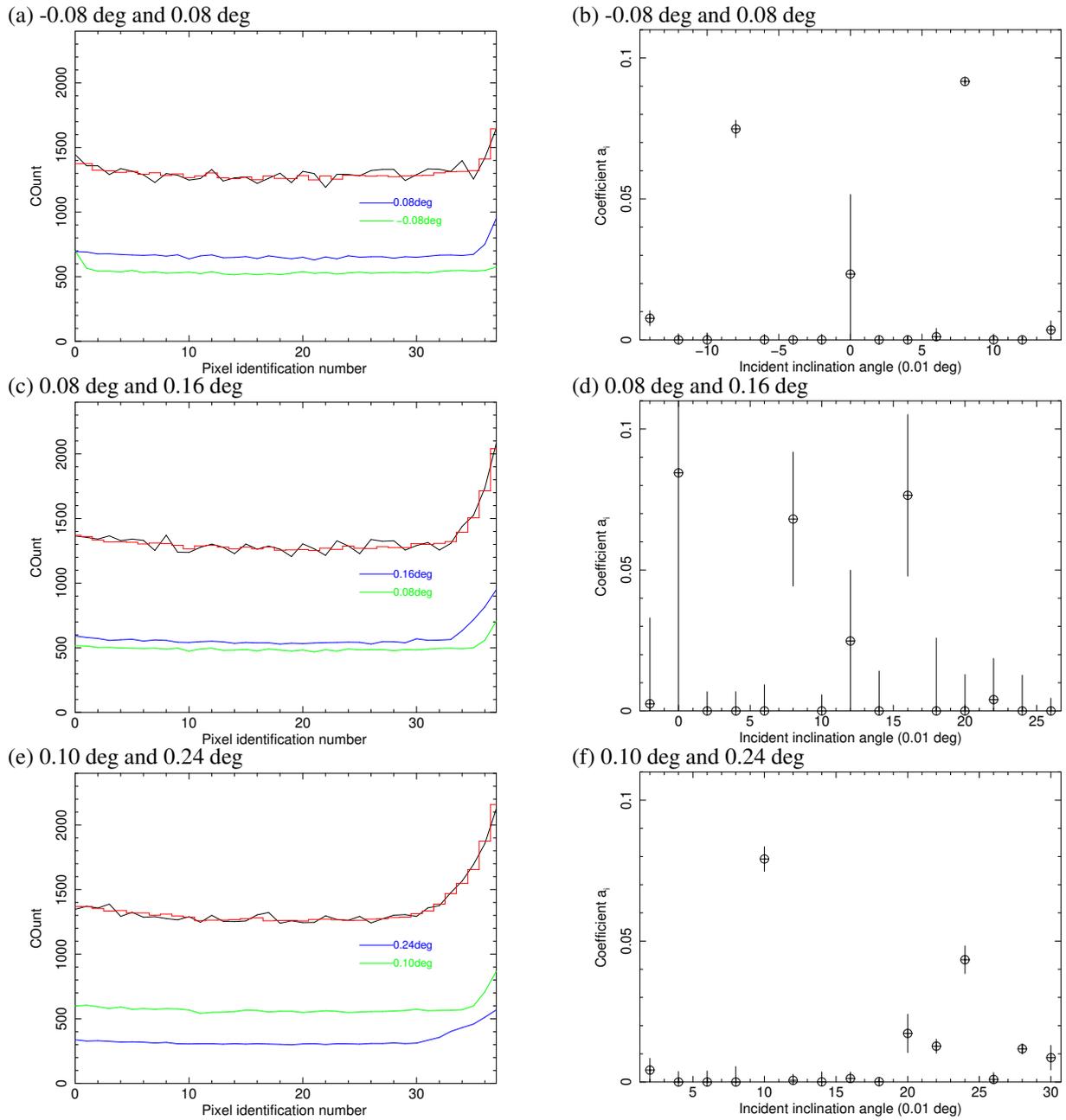

   \vspace*{-2cm}
  \begin{minipage}{0.5\hsize}
  (a) -0.08 deg and 0.08 deg
   \end{minipage}\quad
  \begin{minipage}{0.5\hsize}
  (b) -0.08 deg and 0.08 deg
  \end{minipage}
  \begin{minipage}{0.5\hsize}
   \includegraphics[scale=0.3,angle=-90]{c120m8p8.eps}
   \end{minipage}\quad
   \begin{minipage}{0.5\hsize}
   \includegraphics[scale=0.3,angle=-90]{b120m8p8.eps}
   \end{minipage}
  \begin{minipage}{0.5\hsize}
  (c) 0.08 deg and 0.16 deg
   \end{minipage}\quad
  \begin{minipage}{0.5\hsize}
  (d) 0.08 deg and 0.16 deg
   \end{minipage}
  \begin{minipage}{0.5\hsize}
   \includegraphics[scale=0.3,angle=-90]{c120p8p16.eps}
   \end{minipage}\quad
   \begin{minipage}{0.5\hsize}
   \includegraphics[scale=0.3,angle=-90]{b120p8p16.eps}
   \end{minipage}
  \begin{minipage}{0.5\hsize}
  (e) 0.10 deg and 0.24 deg
   \end{minipage}\quad
  \begin{minipage}{0.5\hsize}
  (f) 0.10 deg and 0.24 deg
   \end{minipage}
  \begin{minipage}{0.5\hsize}
   \includegraphics[scale=0.3,angle=-90]{c120p10p24.eps}
   \end{minipage}\quad
   \begin{minipage}{0.5\hsize}
   \includegraphics[scale=0.3,angle=-90]{b120p10p24.eps}
   \end{minipage}
   \vspace*{2cm}
 \caption{Same as figure \ref{fitting1}. (a)(c)(d) Only sum of 91-100th layers is plotted. Blue and green lines are a model components for smaller and larger incident angle, respectively. }
   \label{fitting2}
\end{figure}

Next, we investigated whether two neighboring gamma-ray sources with a small separation angle can be distinguished using MCCC. The histograms of the two incident angles (each has 0.6M events) were summed and then fitted in the same way as above. Figure \ref{fitting2} showed an example of three cases; a combination of -0.08 degree and 0.08 degree, 0.08 degree and 0.16 degree, and 0.10 degree and 0.24 degree. As can be seen from the figure; two separated peaks appear in the right panels, indicating that two gamma-ray sources can be distinguished for all three cases. From the left panels where blue and green lines are the model components for the smaller and larger incident angles, respectively, it can be seen how the two incident angles contribute to the excess edge profile. Two sources with negative and positive angles (i.e.  located across the center of field-of-view) can be easily distinguished because the edge-peak appears at different sides. On the other hand, two sources with the same sign of angle is more difficult because the edge peak appears at the same side. In this case, the different slope of the peak profiles of the two angles is a key to distinguish them.
  Note that the photon statistics of the excess edge profile becomes larger for the larger inclination angles, and thus the sensitivity could be better for the larger inclination angle.

In summary, MCCC with 1 m column height, 0.5 mm pixel size, 100 layers, and 10 columns for the 1-D direction make it possible to distinguish two sources with a separation of 0.1 degree for 0.6M Compton-reconstructed events. The insensitive edge region of 0.5 mm in pixel sensors make it impossible to determine the incident angle for the incident angle of $<$0.06 degree. A larger pixel size to save power will make the annular resolution worse. More columns with small sensors will improve the sensitivity.

\section{Discussion}

  In this paper, we demonstrated an example of MCCC with 100 layers, 0.5 mm pixel, 10$\times$10 columns, that gives 0.1 degree resolution.
A larger pixel size will reduce the angular resolution; but even in the case of 1 mm pixel size, an expected angular resolution of 0.2 degree is enough better than that of the conventional Compton camera.

  In this simulation, we showed that 0.6 M Compton-reconstructed events provide 0.1 degree resolution. A smaller number of events reduces the angular resolution. Note that this number would not depend on the detector configuration. In this paper, we assumed a rich resource; however, even in the case of less resource (smaller number of layers, no CZT), a longer data-taking time could provide a similar total number of reconstructed events.

  For example, the 511 keV line from the Galactic center is around $1\times10^{-4}$ photons cm$^{-2}$ s$^{-1}$ from the central $\sim1$ degree$^2$ region. The AMEGO-X effective area of $\sim$600 cm$^2$ at 511 keV for the Compton reconstructed events and 0.3 year observation provide 0.6 M events. Therefore, MCCC could provide a meaningful result for bright MeV gamma-ray sources.

  On the other hand, here we did not consider the background signals, such as the in-orbit background, whose estimation in the MeV band is rather complex and not yet established. They would become larger. Exact sensitivity should be estimated by properly modeling those backgrounds and is beyond the scope of this paPer. If we analyze data in a narrow energy band around e.g. 511 keV, the contribution of background events will be small.
  
  For lower energy gamma-rays, the reconstruction efficiency decreases and photo-absorption events which leave a single hit can be used in the same way. However, in this case, the background rejection by Compton reconstruction is not available and thus the coded-mask for photo-absorption events will work better. 
 For higher energy gamma-rays, the reconstruction efficiency also decreases. In this case, the coded-mask also could not work well because the stopping power also decreases and thus does not work as a mask.

The rwo-dimensional cases could be demonstrated in the similar way as shown above. We assumed a gamma-ray source at infinity distance for astrophysical observation. However, MCCC can be applied to the medical case, such as PET, where the gamma-ray source distance is not infinite. Even in that case, the somewhat different shadow effect could appear in the count distribution, and it also provides angular resolution. We could perform gamma-ray imaging with much better angular resolution than that of the Compton camera which is widely used for PET.

\section{Summary and Conclusion}

We proposed a new concept of MCCC and demonstrated that the angular resolution can be less than 0.1 degree. This is much better than the conventional Compton camera. The recently developed AstroPix sensor is a good candidate to construct the MCCC. The presented algorithm obtaining the direction of injected gamma-rays is not sophisticated and can be improved. MCCC needs good statistics as much as possible for making the edge component clearer in the count distribution of the first scattering points, and thereby we prefer to have columns and layers as much as possible.

\newpage

\end{document}